\documentclass[fleqn]{elsart}

\usepackage[OT2,OT1]{fontenc}
\newcommand\cyr{%
\renewcommand\rmdefault{wncyr}%
\renewcommand\sfdefault{wncyss}%
\renewcommand\encodingdefault{OT2}%
\normalfont
\selectfont}
\DeclareTextFontCommand{\textcyr}{\cyr}
\usepackage{amsmath}
\usepackage{amsfonts}
\usepackage{amssymb}
\usepackage{graphicx}
\usepackage{bm}

\def\be{\begin{equation}}
\def\ee{\end{equation}}
\def\ba{\begin{eqnarray}}
\def\ea{\end{eqnarray}}
\def\bs{\begin{subequations}}
\def\es{\end{subequations}}

\def\rme{\text{e}}
\def\rmd{\text{d}}
\def\rmi{\text{i}}
\def\p{\partial}

\def\B{\Box}
\def\a{\alpha}
\def\b{\beta}
\def\s{\sigma}

\def\g{\gamma}
\def\t{\theta}
\def\k{\kappa}
\def\tk{\tilde\kappa}
\def\bk{\bar\kappa}
\def\erf{{\rm erf}}
\newcommand{\Eq}[1]{(\ref{#1})}

\begin{document}

\begin{frontmatter}

\title{Kinks of open superstring field theory}

\author{Gianluca Calcagni},
\ead{gianluca@gravity.psu.edu}
\address{Institute for Gravitation and the Cosmos, Department of Physics,\\ The Pennsylvania State University,
104 Davey Lab, University Park, PA 16802, United States}
\author{Giuseppe Nardelli},
\ead{nardelli@dmf.unicatt.it}
\address{Dipartimento di Matematica e Fisica, Universit\`a Cattolica,
via Musei 41, 25121 Brescia, Italia}
\address{INFN Gruppo Collegato di Trento, Universit\`a di Trento, 38100 Povo (Trento), Italia}

\begin{abstract}
We construct approximate kink solutions of supersymmetric open string field theory at lowest level when non-local operators in the tachyon effective action are fully taken into account. To this purpose we derive two duplication formul\ae\ for products of incomplete gamma functions, which determine the level of approximation of the solutions. The time kink is an instanton of the Euclidean theory with a well-defined tunneling probability. The spatial kink solution represents an unstable D9-brane decaying into a stable D8-brane. We calculate the ratio of the brane tensions, which almost exactly reproduces the expected value.
\end{abstract}



\end{frontmatter}


\section{Introduction}

The prototype of instanton in local scalar theories is the classical Euclidean solution for a double-well potential, $-V(\Phi)\sim \Phi^2-\Phi^4$, representing a scalar field rolling between the local maxima of $-V$. The study of  instantonic solutions is an essential tool to understand the vacuum structure of the corresponding Lorentzian theory with the potential upside down. In particular, one can understand whether (i) the ground state is degenerate or not, (ii) tunnel effects can take place between the two localized minima of the potential $V$, (iii) such a penetration through the barrier removes the degeneracy of the ground state and what is the splitting, and (iv) the system oscillates between the two minima and what is the frequency of the resonant state \cite{Col88,Raj89,KST}.

The same problem is neither trivial nor of mere academic interest as far as non-local theories are concerned. In fact, the simplest example of a non-local scalar with a static ({\it i.e.}, zero-momentum) double-well potential is provided by the tachyon of open string field theory (OSFT, reviewed in \cite{ohm01,ABGKM,sen04,FK}). 
In the case of nonchiral bilocal supersymmetric OSFT \cite{PTY,AKBM,AJK}, the Euclidean $(1/2,1)$-level effective equation for a homogeneous tachyon is (up to a field redefinition)
\be\label{exa}
(\p_t^2-m^2)\rme^{-s\p_t^2}\Phi=\s\Phi\,\rme^{s\p_t^2}\Phi^2\,,
\ee
where $t$ is Wick-rotated time and $m^2<0$, $s>0$ and $\s>0$ are parameters fixed by the theory. In the appendix we briefly review how to obtain this equation from the fundamental OSFT action. Solving Eq.~\Eq{exa}, even numerically, has been proved to be a formidable task and a simplified version of the dynamics was proposed in \cite{AJK}, where the potential is purely local:
\be\label{a1}
(\p_t^2-m^2)\rme^{-s\p_t^2}\Phi=\s\Phi^3\,.
\ee
While for local systems a classical rolling solution between two maxima is a textbook exercise for the double-well potential, non-local dynamics can (and in fact does) spoil the neat correspondence between classical rolling and kink solutions. In fact, there exist Lorentzian solutions which start from a local maximum and apparently `climb' the potential up and down indefinitely \cite{roll}. The same phenomenon was previously noted also for the bosonic cubic potential \cite{MZ,FGN,CST,KORZ,BMNR}.\footnote{Analytic background-independent solutions for marginal deformations of Berkovits' string field theory were found in \cite{Er07,Ok07a,Ok07b,KO}.}

For certain values of the parameters, a kink solution of Eq.~\Eq{a1} has long been established numerically \cite{vol03} (also on a cosmological background \cite{Jou07,AKV2}) and by an existence theorem \cite{pro06}.\footnote{Kink solutions of the $p$-adic equation are discussed in \cite{MZ,vol03,BFOW,vla05,VV,Jo072}; existence (and uniqueness) theorems can be found in \cite{VV,Jo072,vla06}.}\footnote{The existence of a kink was also argued in \cite{are04} on Minkowski and in \cite{AJ,AKV1} on a cosmological background. However, the action was truncated at second order in time derivatives and, due to the fundamental difference between non-local and finite-order theories, this type of evidence could not be considered as conclusive.} On the other hand, little is known about the solutions of Eq.~\Eq{exa}. The first goal of this paper is to find an analytic, bounded, (very-well) approximated instanton solution of Eq.~\Eq{a1}. In doing so we shall provide a method which is viable also for the original system Eq.~\Eq{exa}. As a byproduct, we are able to verify quantitatively how well the approximation of \cite{AJK} works for kink solutions. In addition, all the basic physical quantities associated with the instanton solutions in the local double-well potential can be reproduced also in this non-local context: energy splitting, penetration barrier, resonance frequency and so on. 

Moreover, the time-independent kink profile ({\it i.e.}, where $t$ is a spatial coordinate) is a Lorentzian soliton describing the decay of an unstable D$p$-brane into a stable D$(p-1)$-brane. This is Sen's first and second conjectures, which we verify explicitly by calculating the ratio of the brane tensions. The result is very close ($\sim 99.8\%$) to the theoretical value $\tilde{\cal T}_{p-1}/{\cal T}_p=\sqrt{2}\pi$ (in $\alpha'$ units; we will explain this `descent relation' \cite{sen04} in Section \ref{tension}). This datum adds to the evidence that the diffusion equation is a natural consequence of the conformal symmetry of string theory, as will be discussed in \cite{cust}.

In \cite{cuta4} we constructed an explicit, global, approximate solution of the effective equation of motion \Eq{a1}. We made use of the diffusion equation approach developed in \cite{roll,cuta2,cuta3} (see also \cite{FGN,vol03,Jou07,vla05,Jo081,Jo082,MuN} in relation with the same method and \cite{BK1,BK2} for another recent treatment of non-local scalars). Non-local systems are reinterpreted as living at the fixed point of a higher-dimensional flow along the diffusion time $r$. The scalar $\Phi(t,r)$ obeys a heat equation which localizes all non-local operators, so that they act as shifts along $r$. Choosing a sign distribution as initial condition for the diffusing system, the diffusion equation evolves it to the error function 
\be\label{erf}
\erf\left(\frac{t}{\sqrt{4r}}\right) \equiv \frac{2}{\sqrt{\pi}} \int_0^{t/\sqrt{4r}} \rmd \tau\,\rme^{-\tau^2}\,,
\ee
which solves Eq.~\Eq{a1} with very good accuracy.

The problem in this and other non-local systems is that, in general, powers of the candidate solution do not obey the diffusion equation and the potential does not translate naturally along $r$. This is true also for the error function. Its validity and the level of approximation were checked  locally via an asymptotic expansion and a matching of the leading coefficients \cite{cuta4}. Here we propose an analytic method which allows one not only  to solve globally the same problem in a more efficient way, but also to face the much more difficult problem of nested non-local operators, as in Eq.~\Eq{exa}.

As a first step, which is also of mathematical interest, we shall derive an approximate formula expressing products of incomplete gamma functions $\g(\a,z^2)$ $\times\g(\b,z^2)\dots$ as an incomplete $\g$ with rescaled arguments. The error function is the special case $\a=1/2$, which is deformed to general incomplete $\gamma$'s by non-local operators.

While in previous works approximate solutions were seeked for particular problems, here we operate the other way round and ask to which systems, through the approximate scaling formula, the incomplete gamma function is an analytic global solution. The answer not only includes the instantonic solution\footnote{When considering homogeneous configurations, the spatial volume is factorized from the Euclidean action and the following kink-type solutions are `quantum-mechanical' instantons.} of Eq.~\Eq{a1}, as expected from \cite{cuta4}, but also the highly non-local Eq.~\Eq{exa}, previously inaccessible by analytical methods, and two cosmological solutions for, respectively, a cubic and exponential potential. We shall report on the cosmological cases in a companion paper \cite{cuta6}. 

The paper is organized as follows. In Section \ref{dup} we derive two approximate scaling formul\ae\ for incomplete gamma functions and discuss their relevance for diffusing systems. The instanton of susy OSFT for the approximate  potential and its associated tunneling probability are discussed in Section \ref{tuna}, while in Section \ref{tension} we compute the brane tension of the Lorentzian soliton. Section \ref{exsu} deals with the exact non-local potential. The main results are summarized and discussed in Section \ref{conc}. We shall work in units of the Regge slope, $\a'=1$.


\section{Duplication formul\ae\ and diffusion equation}\label{dup}

We shall be interested in non-local systems described by the scalar field
\be\label{phi}
\Phi=\Phi(\a,z)=\frac{\g\left(\a,z^2\right)}{\Gamma(\a)}\,,
\ee
where $\g$ is the lower incomplete gamma function (see Section 8.35 of \cite{GR}):
\be
\gamma(\a,z^2)=2\int_0^z \rmd \tau\, \tau^{2\a-1} \rme^{-\tau^2}\,.
\ee
Here $z$ is an evolution variable we shall keep positive for simplicity. $z=0$ is a branch point for $\Phi$ except when $\alpha$ is half-integer or integer, where $\Phi$ admits the odd (or even) analytic continuation to the whole real axis. In such cases $\Phi$ is analytic at the origin. More explicitly, for any non-negative integer $N$
\ba
\Phi\left(N+\frac12, z\right)&=& {\rm erf} (z)- \rme^{-z^2}\sum_{j=0}^{N-1}\frac{z^{2 j+1}}{\Gamma(j+3/2)}\,,\\
\Phi(N+1, z)&=& 1- \rme^{-z^2}\sum_{j=0}^{N}\frac{z^{2 j}}{j!}\,,
\ea
where the first sum is omitted when $N=0$. Some properties we will make use of are the asymptotic expansions
\ba
\Phi(\a,z) &\ \stackrel{z\to 0}{\sim}\ & \frac{z^{2\a}}{\Gamma(\a+1)}\,,\label{asy0}\\
\Phi(\a,z) &\ \stackrel{z\to \infty}{\sim}\ & 1\,,\label{asyinf}
\ea
the shift property
\be
\Phi(\a+1, z) = \Phi(\a,z) - \frac{z^{2 \a} \rme^{-z^2}}{\Gamma(\a+1)}\,,\label{prop1}
\ee
and
\ba
\p_z\Phi(\a,z) &=& \frac{2 z^{2\a-1}\rme^{-z^2}}{\Gamma(\a)}\,,\label{der2}\\
\p^2_z\Phi(\a,z)&=& \p_z\Phi(\a,z) \left(\frac{2\a-1}{z}-2z\right)\,.\label{der12}
\ea


\subsection{First duplication formula}\label{sca1}

The problem is to conveniently rewrite powers of $\Phi$ in a form which manifestly obeys the diffusion equation. We begin with a product of two $\Phi$'s with different first argument, which is a double integral:
\be
\Phi(\a, z)  \Phi(\beta, z)= \frac{4}{\Gamma(\a)\Gamma(\beta)}\int_0^z \rmd \tau_1 \rmd \tau_2\, \tau_1^{2\a-1} \tau_2^{2\beta-1}\rme^{-(\tau_1^2+\tau_2^2)}\,.
\ee
In polar coordinates $(\varrho,\t)$ the integrand function factors into a radial and an angular part. This integral is not easier to perform, since the integration domain, a square of side $z$ with one corner at the origin of the $(\tau_1,\tau_2)$ plane, is not isotropic. A useful approximation, adopted also in Debye's model of phonons, consists in replacing the square with the quarter of a disk centered at the origin and with radius $R=\bk z$, where $1<\bk<\sqrt{2}$. The first quadrant is $\varrho\in[0,R]$, $\t\in[0,\pi/2]$. Then,
\ba
\Phi(\a, z)  \Phi(\beta, z) &\approx& \frac{4}{\Gamma(\a)\Gamma(\beta)}\int_0^{\pi/2} \rmd \theta (\sin\theta)^{2\a -1} (\cos\theta)^{2\b -1}\int_0^R \rmd \varrho \varrho^{2(\a+\b)-1}\rme^{-\varrho^2}\nonumber\\
&=& \frac{1}{\Gamma(\a+\b)} \gamma(\a+\b, R^2)\,,\nonumber
\ea
giving the first duplication formula of this paper:
\be\label{dupli1}
\Phi(\a, z)  \Phi(\beta, z) \approx \Phi(\a+\b, \bk z)\,.
\ee
The constant $\bk$ can be determined by choosing $R$ so that to preserve the isotropic integration area, $z^2= \pi R^2/4$. Hence $\bk =2/\sqrt{\pi}\approx 1.128$ for any $\a$ and $\b$ (half-integer if Eq.~\Eq{dupli1} is to be considered on the whole real axis). Actually, a more accurate estimate of $\bk$ is obtained by matching the behaviour of Eq.~\Eq{dupli1} at the origin, yielding
\be\label{kapp1}
\bk=\left[\frac{\Gamma(\a+\b+1)}{\Gamma(\a+1)\Gamma(\b+1)}\right]^{\frac{1}{2(\a+\b)}}\,.
\ee
One can check that Eq.~\Eq{dupli1} is satisfied with good accuracy when taking Eq.~\Eq{kapp1}. For instance, one can estimate the quantity
\be\label{DX}
\Delta_{\rm max}\equiv \mathop{\rm sup}_{z}\Delta(z)\equiv \mathop{\rm sup}_{z}\left|\frac{{\rm LHS}-{\rm RHS}}{{\rm LHS}+{\rm RHS}}\right|\,,
\ee
where LHS and RHS are, respectively, the left- and right-hand side of Eq.~\Eq{dupli1}. Since we shall make use of the first duplication formula with $\a=1/2$ and $\b=1$, let us take these values. For the rescaling coefficient Eq.~\Eq{kapp1}, 
\be\label{bk}
\bk =\left(\frac32\right)^{1/3}\approx 1.145\,,
\ee
and $\Delta_{\rm max}\approx 1.3\%$, while using the isotropic value $\bk=2/\sqrt{\pi}$ the maximum error is above $2\%$. From now on we shall ignore the latter.

\begin{figure}
\begin{center}
\includegraphics[width=8cm]{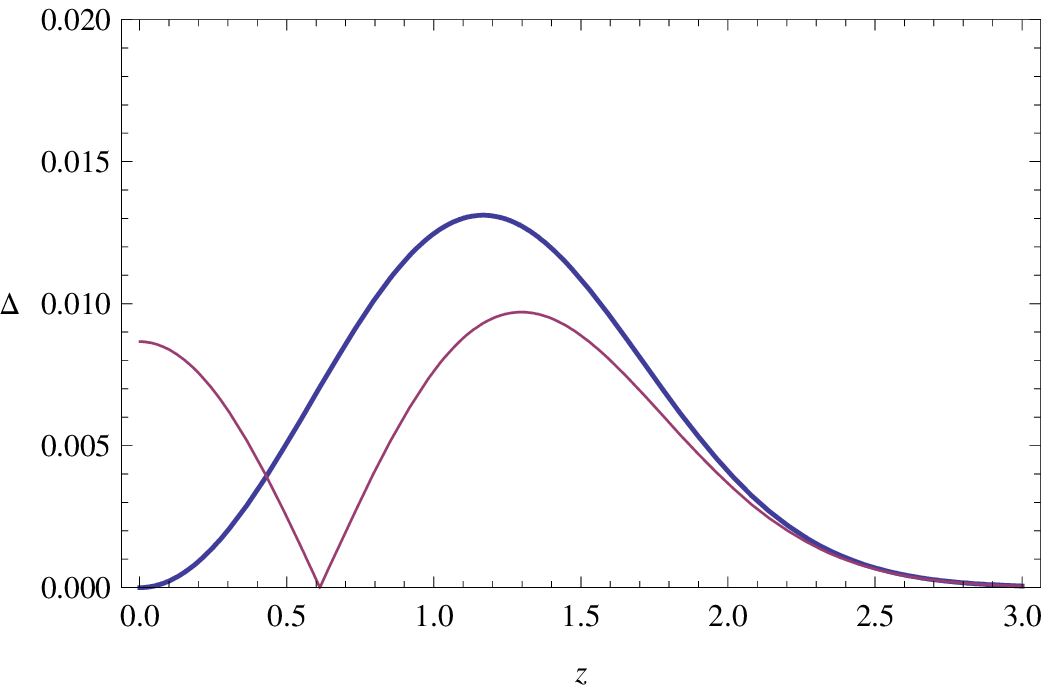}
\end{center}
\caption{\label{fig1} $\Delta$ of the first duplication formula, Eq.~\Eq{dupli1}, with $\bk=(3/2)^{1/3}$ (theoretical value, thick line) and $\bk=\bk_{\rm num}\approx 1.138$ (numerical value, thin line). Here $(\a,\b)=(1/2,1)$.}
\end{figure}

At this point a numerical method to improve the result is ready at hand. In fact, one can find the value of $\bk$ minimizing the global error:
\be\label{del}
\delta\equiv \mathop{\rm inf}_{\bk} \sqrt{\int_0^{\bar{z}} \rmd z \Delta^2(z)}\,,
\ee
where $\bar{z}=O(10)$ is sufficient. We get $\delta\approx 1.0\%$ and $\Delta_{\rm max}\approx 0.97\%$ at $\bk_{\rm num}\approx 1.138$, which is very close to the theoretical value (see Fig.~\ref{fig1}).

An alternative criterion is to minimize the $L^2$ norm of Eq.~\Eq{dupli1} and integrate only $({\rm LHS}-{\rm RHS})^2$ in Eq.~\Eq{del}. In general this yields a higher $\Delta_{\rm max}$ but a similar result as far as the global accuracy of the solution is concerned.

Since non-locality has been all encoded in argument rescalings of the scalar field, we will have to deal only with \emph{algebraic} equations. Therefore Eq.~\Eq{DX} or \Eq{del} or the $L^2$ norm are sufficient to check the accuracy of the equations. Nevertheless, one can show that Sobolev-type norms of the form, e.g., $W_n=\int \rmd z [{\rm LHS}^{(n)}-{\rm RHS}^{(n)}]^2$, where $(n)$ denotes the $n$th derivative with respect to $z$, recover about the same values of $\kappa$ obtained with Eqs.~\Eq{DX} and \Eq{del}. For instance, we have checked that the first Sobolev norms $W_1$, $W_2$, \dots for Eq.~\Eq{dupli1} yield $\kappa$'s between the isotropic value $1.128$ and Eq.~\Eq{bk}, with errors comparable with those above. This indicates that the estimate of the error is reasonably `stable' with respect to derivatives.


\subsection{Second duplication formula}\label{sca2}

\begin{figure}
\begin{center}
\includegraphics[width=8cm]{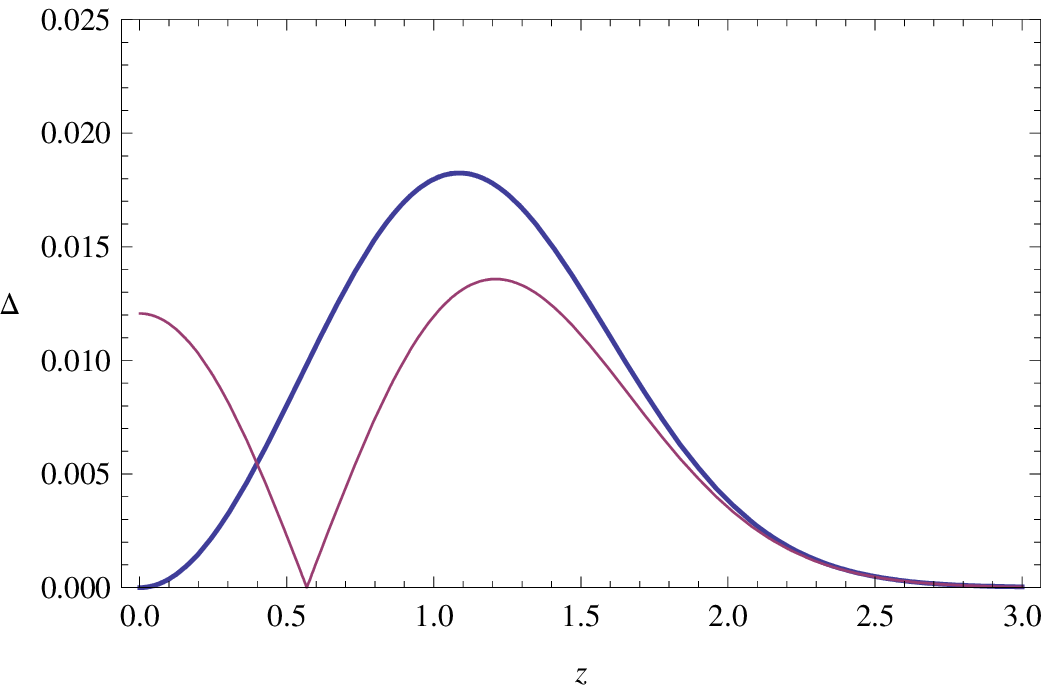}
\end{center}
\caption{\label{fig2} $\Delta$ with the theoretical (thick line) and numerical (thin line) value of $\kappa$. Here $(n,\a)=(3,1/2)$.}
\end{figure}

Taking $\a=\b$ and iterating Eq.~\Eq{dupli1}, one obtains the scaling formula
\be\label{dupli2}
[\Phi(\a,z)]^n \approx \Phi(n\a,\kappa z)\,,
\ee
where
\be
\k=\kappa(n,\a)\equiv \frac{\left[\Gamma(n\a+1)\right]^{1/{(2n\a)}}}{\left[\Gamma(\a+1)\right]^{1/{(2\a)}}}\,.\label{kappa2}
\ee
The value $\a=1/2$ is of particular interest, because it corresponds to the kink of open string field theory:
\be
\Phi\left(\frac12,z\right)=\erf(z)\,.
\ee
In this case,
\be\label{kapp}
\kappa = \frac{2}{\sqrt{\pi}} \left[\Gamma\left(\frac{n}{2}+1\right)\right]^{1/n}\,.
\ee
We can check that the error propagating from Eq.~\Eq{dupli1} stays very low. When $(n,\a)=(3,1/2)$, the maximum error is $\Delta_{\rm max}\approx 1.8\%$, which is rather good considering we have resorted only to analytic tools. The minimum $\delta$ is $\delta\approx 1.3\%$ at $\kappa\approx 1.23$, very close to the theoretical value $\kappa=(6/\pi)^{1/3}\approx 1.24$. The maximum error is $\Delta_{\rm max}\approx 1.4\%$, slightly better than the one obtained by the analytic approximation (see Fig.~\ref{fig2}).

For the particular case $n=1+1/\a$, the sup-norm of $\Delta$ depends on $\a$ only very mildy. In fact, a change in $\a$ results only in a shift of the position of the maximum to slightly higher $z$, while its height is almost unaffected. For this reason Fig.~\ref{fig2} is a reference also for other values of $\a$ when $n=1+1/\a$.

An interesting possibility is the analytic continuation to real values of $n$. The scaling formula Eq.~\Eq{dupli2} is still a good approximation, with errors of the same order of magnitude of the ones previously discussed. Finally, it is worth mentioning that  Eq.~\Eq{dupli2} seems to work well even for negative values of $n$, proving it is rather `stable' against the variation of its parameters. We are confident that the range of applications of both duplication formul\ae\ will not be exhausted within this paper.


\subsection{Diffusion equation}

Let us consider a cosmological-type setting where all fields are homogeneous and live in a $D$-dimensional universe characterized by a Hubble parameter $H(t)$, where $t$ is proper time. Defining
\be
t\equiv 2 \sqrt{r}\,z
\ee
and the d'Alembertian operator (on scalar fields; metric signature ${-},{+},\cdots,{+}$) $\B=-\p_t^2-(D-1) H\p_t$, the incomplete gamma function obeys the diffusion equation
\be\label{difeq}
\left(\B+\p_r\right)\,\Phi\left(\a,\frac{t}{2\sqrt{r}}\right)=0\,,
\ee
provided
\be\label{hubb}
H=\frac{1-2\a}{(D-1)}\frac1t\,.
\ee
One gets Minkowski spacetime when $\a=1/2$. For any $s$,
\be\label{tra}
\rme^{s\B}\Phi\left(\a, \frac{t}{2\sqrt{r}}\right)=\rme^{-s\p_r}\Phi\left(\a,\frac{t}{2\sqrt{r}}\right)=\Phi\left(\a,\frac{t}{2\sqrt{r-s}}\right)\,.
\ee
The action of an exponential non-local operator on $\Phi$ is that of a simple shift along the extra direction $r$, but only if the d'Alembertian and derivatives of $r$ commute, $[\B,\p_r]\Phi=0$. This condition can be imposed only if the Hubble parameter does not depend on $r$.

Comparing Eqs.~\Eq{der12} and \Eq{difeq}, we infer an identity involving only first-order derivatives: $(2r\p_r+t\p_t)\Phi=0$.

We conclude this section with another formula we shall often employ. On a flat background, $\Phi(1,z)=1-\rme^{-z^2}$ does not obey the diffusion equation. However, one can show that
\be\label{gf1}
\rme^{\b\p_z^2}\Phi(1,z)=1-\frac{\rme^{-z^2/(1+4\b)}}{\sqrt{1+4\b}}\,.
\ee
To get this result, one can use the series representation of the non-local operator and recognize the generating function of even Hermite polynomials \cite{Foa81,GJ}. A simpler alternative is to Fourier transform $\Phi$.


\section{Supersymmetric OSFT: Approximate potential}\label{apsu}

We are ready to write a non-local Klein--Gordon equation for the scalar field given in Eq.~\Eq{phi}. We start with the simplest case where non-locality is confined to the kinetic term. The normalization of $\Phi$ is fixed to 1 without loss of generality; this will only affect the normalization $\s$ of the self-interacting potential. The mass $m$ of the field is tuned by $r$.

Recasting the $r$ derivative in the diffusion equation \Eq{difeq} in terms of Eq.~\Eq{der2} and making use of Eq.~\Eq{prop1}, one has
\be\label{use}
\left(\B-\frac{\a}r\right)\Phi(\a,z)=-\frac{\a}r\Phi(\a+1,z)\,.
\ee
Let us define
\ba
&&\tilde\Phi \equiv \rme^{s\B}\Phi=\Phi(\a,\kappa z)\,,\\
&&m^2\equiv -\frac{\a\kappa^2}{r}\,,\qquad \sigma=-m^2\,,
\ea
where
\ba
s &=&\frac{\kappa^2-1}{\kappa^2}r\,,\label{s}\\
\kappa&=& \kappa\left(1+\frac1\a,\a\right)\,.
\ea
Taking the scaling formula \Eq{dupli2} for $n=1+1/\a$, Eq.~\Eq{use} for $r\to r/\kappa^2$ becomes
\be\label{geom}
\left(\B+m^2\right)\tilde\Phi\approx-\s \Phi^{1+1/\a}\,.
\ee
For natural $n$ and $\a$ integer or half integer (solutions analytic at the origin), only $(n,\a)=(2,1)$ or $(n,\a)=(3,1/2)$ are possible.

While in \cite{roll,cuta4,cuta2,cuta3} the local limit $\Phi=\tilde\Phi$ corresponded to a local system with the same type of potential, here it gives rise to an altogether different theory. This is because we are in the context of the inverse problem, where the dynamics is defined through the field and, in turn, the local limit of $\Phi$ is trivial.

The local limit corresponds to $s\to 0$, {\it i.e.}, $\kappa\to1$ and, from the scaling formula, $n\to 1$ (and then $\a$ large). The right-hand side of Eq.~\Eq{phi} tends to zero at large $\a$, in agreement with Eq.~\Eq{geom} in the same limit, $\Phi(\a,z)\sim\Phi(\a+1,z)$, which admits $\Phi=0$ as the only solution. Conversely, when $m^2$, $\s$ and $n$ are kept fixed Eq.~\Eq{geom} formally reduces to the properly said \emph{local} system.

\begin{figure}
\begin{center}
\includegraphics[width=8cm]{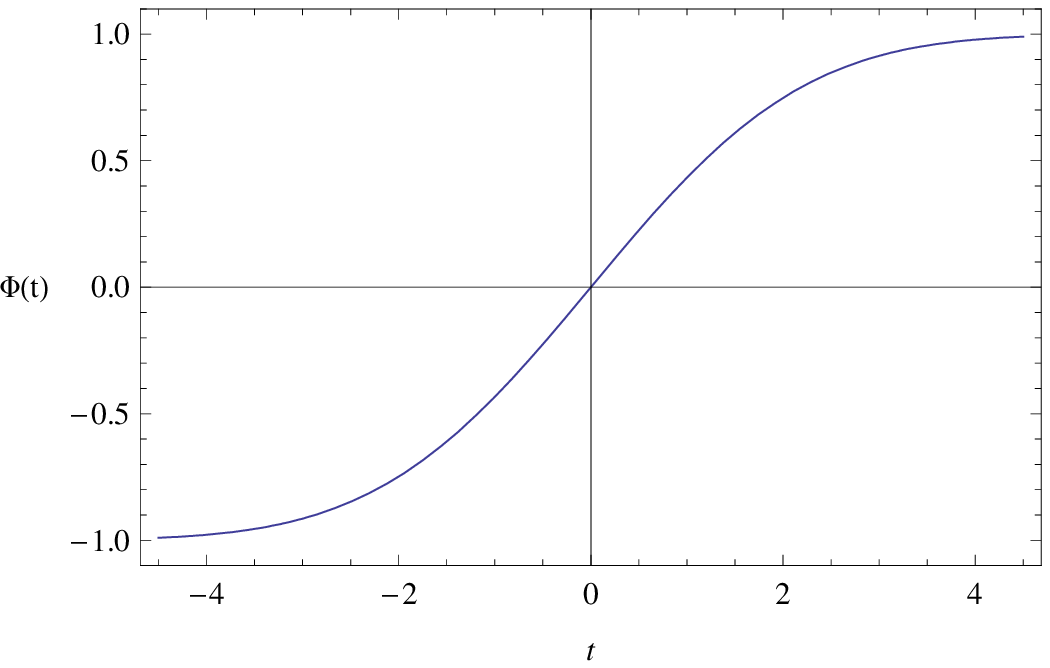}
\includegraphics[width=8cm]{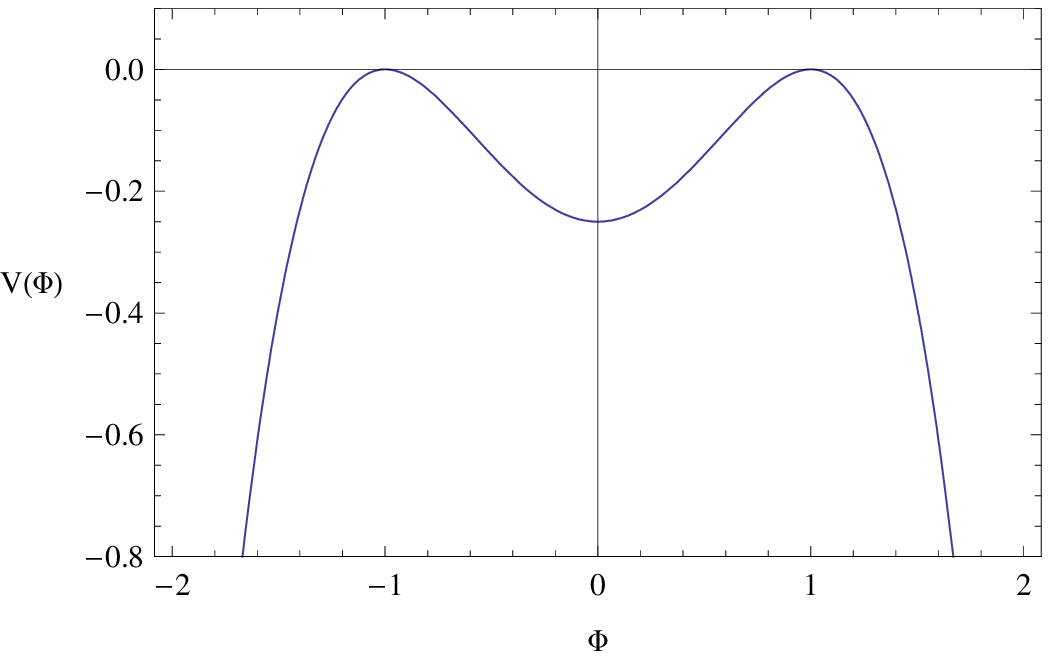}
\end{center}
\caption{\label{fig3} Non-local instanton (upper panel) and its static potential $V=\Phi^2/2-\Phi^4/4-1/4$ (lower panel), where $n=3$, $\a=1/2$ and $r=1+s_*\approx 1.523$.}
\end{figure}

The case $(n,\a)=(3,1/2)$ corresponds to the effective \emph{Euclidean} equation of the supersymmetric tachyon at lowest level in the field expansion and where non-locality is transferred completely onto the kinetic term.
For a homogeneous field and $\B=-\p_t^2$, Eq.~\Eq{geom} becomes 
\be\label{susya}
(\p_t^2-m^2)\rme^{-s\p_t^2}\Phi=\s\Phi^3\,.
\ee
Since $r$ is positive, Eq.~\Eq{susya} reproduces Eq.~\Eq{a1} as promised, where $t$ plays the role of Wick-rotated time. Therefore, this kink-type solution can be regarded as an instanton (see Fig.~\ref{fig3}).

Let us fix $r$ by requiring the mass to be the one of the string tachyon in units of the Regge slope: $m^2=-1/2$. Hence, $r=\kappa^2$. Then, $s=\kappa^2-1\approx 0.539$ for the theoretical value of $\kappa$, while 
\be\label{esse}
s\approx 0.515
\ee
for the numerical estimate of $\kappa$. This is in excellent agreement with the value of $s$ dictated by conformal symmetry in string field theory,
\be\label{stv}
s=s_*=2 \ln (3^{3/2}/4)\approx 0.523\,.
\ee
We will come back to this point in the conclusions.

Since non-locality affects the kinetic and potential energy as a relative rescaling of the coordinate, the usual asymptotic properties of a scalar field are automatically recovered. Namely, at $t=\pm\infty$ the effective potential dominates over the kinetic term and the scalar pressure is negative ($P+E\sim0$, effective `cosmological constant'), while near the origin $t\sim 0$ the pressure is equal to the energy ($P\sim E$, stiff matter).


\subsection{Tunneling}\label{tuna}

In this subsection we shall reproduce all the basic features associated with the non-local instanton $\Phi(z)$.  Its Euclidean action (per unit spatial volume) is
\be\label{euca}
S_{\rm E}=\frac12\int \rmd t \left[\Phi(\B+m^2)\tilde\Phi+\frac{\s}2\Phi^4+ \frac{\sigma}{2}\right]\,,
\ee
where $\B=-\p_t^2$ and  the constant $\sigma/4$ has been added to  set the zero energy of the potential at the minima $\Phi=\pm1$. Variation of this action with respect to $\Phi$ yields the equation of motion \Eq{susya}.
Pulling back a perturbed classical solution $\Phi=\Phi_{\rm cl}+\delta\Phi$ into the action, where $\delta\Phi$ is a quantum fluctuation, one can expand Eq.~\Eq{euca} as
\be
\label{azi}
S_{\rm E}=S_0[\Phi_{\rm cl}]+\frac12\int  \rmd t \delta\Phi(t)\left[\frac{\delta^2 S_{\rm E}}{\delta\Phi(t)\delta\Phi(t)}\right]_{\Phi=\Phi_{\rm cl}}\delta\Phi(t)+\dots\,.
\ee
Two particular cases are of interest. The first corresponds to the trivial solution $\Phi_{\rm cl}\equiv \pm1$ describing the (Lorentzian) particle localized in one of the minima of the potential. Then, $S_0[\pm 1]=0$ whereas the quantum fluctuation operator reads
\be
\label{fr}
\left(\frac{\delta^2 S_{\rm E}}{\delta\Phi^2}\right)_{\Phi_{\rm cl}=\pm1}= \rme^{s\B}\left(\B + m^2 + 3 \s\right)\equiv\rme^{s\B}\left(-\p_t^2 +\omega^2\right)\,,
\ee
where $\omega^2= m^2+3\s$. If the particle were localized in one of the minima, $\omega$ would be the  proper frequency of the oscillator  and $\omega/2$ the energy of the vacuum state ($\hbar=1$).

The second case is when the classical solution is the instanton given in Eq.~\Eq{erf}, $\Phi_{\rm cl}=\Phi_{\rm inst}$. Such solution describes a (Euclidean) particle rolling between the two maxima of the inverted potential and corresponds, in Lorentzian signature, to the tunneling  between the two vacua. Then, $S_0[\Phi_{\rm inst}]=E_K+V$ can be computed and the integral showed to converge. The kinetic term is (integration on $\mathbb{R}$)
\ba
E_K &=&-\frac12\int \rmd t\, \Phi_{\rm inst}\p_t^2\tilde\Phi_{\rm inst}\nonumber\\
  &=&-\frac{1}{4\sqrt{r}}\int \rmd z\, \Phi_{\rm inst}(z)\p_z^2\Phi_{\rm inst}(\kappa z)\nonumber\\
  &=&\frac{\kappa}{\sqrt{\pi r(1+\kappa^2)}}\,,
\ea
while the potential is ($m^2=-\sigma$)
\ba\label{v}
V &=&\frac{\s}{4}\int \rmd t \left(1-2\Phi_{\rm inst}\tilde\Phi_{\rm inst}+\Phi^4_{\rm inst}\right)\nonumber\\
  &=&\frac{\s\sqrt{r}}{2}\int\rmd z\left\{\left[\Phi_{\rm inst}^2(z)-1\right]^2 + 2\Phi_{\rm inst}(z) \left[\Phi_{\rm inst}(z)-\Phi_{\rm inst}(\kappa z)\right]\right\}\nonumber\\
  &=&\frac{\s\sqrt{r}}{2}\left[N + \frac{4}{\kappa \sqrt{\pi}} \left(\sqrt{1+\kappa^2}-\sqrt{2\kappa^2}\right)\right]\,.
\ea
Here $N$ is the numerical factor
\be
\int\rmd  z \left[\Phi_{\rm inst}^2(z) -1 \right]^2 \approx 1.1201\,,
\ee
which may also be estimated  analytically by using the scaling formula for $\a=1/2$ and $n=4$ and $n=2$:
\ba
\int\rmd  z \left[\Phi_{\rm inst}^2(z) -1 \right]^2 &\approx & \int \rmd z \left[ \gamma \left(2,\frac{2^{5/2}z^2}{\pi} \right)-2  \gamma \left(1,\frac{4 z^2}{\pi} \right) +1 \right] \nonumber\\ &=&\pi -\frac{3\pi}{2^{9/4}}\approx 1.1603\,.
\ea
As expected, the error on the `exact' result (which we shall use from now on) is of order $1\%$. Plugging in the values 
\be\label{val}
r=\kappa^2=1+s_*\approx 1.523\,,
\ee
as dictated by string theory, one gets $S_0\approx 0.612$. This fixes the first term in Eq.~\Eq{azi}. Concerning the second term, we have the quantum fluctuation operator for the instanton
\be
\label{fin}
\left(\frac{\delta^2 S_{\rm E}}{\delta\Phi^2}\right)_{\Phi=\Phi_{\rm inst}}= \rme^{s\B}\left[\B + m^2 + 3 \s \rme^{-s\B} \Phi^2_{\rm inst}\right]\,.
\ee
The last (non-local) term in Eq.~\Eq{fin} can be calculated with a further use of the second duplication formula, Eqs.~\Eq{dupli2} and \Eq{kappa2} with $n=2$ and $\alpha=1/2$, together with Eq.~\Eq{gf1}:
\ba
\rme^{s\p_t^2}\Phi^2_{\rm inst}(z)&\approx&\rme^{s\p_t^2}\gamma\left(1,\frac{t^2}{\pi r}\right)\nonumber\\
&=&1-\frac{\sqrt{\pi r}\,\rme^{-\frac{t^2}{\pi r+4s}}}{\sqrt{\pi r+4s}}\,.
\ea
Up to a non-local factor, the quantum fluctuation operator in the instanton background, Eq.~\Eq{fin}, simply becomes the Schr\"odinger operator with an attractive Gaussian potential:
\be
\label{finn}
\left(\frac{\delta^2 S_{\rm E}}{\delta\Phi^2}\right)_{\Phi=\Phi_{\rm inst}}= \rme^{s\B}\left[ -\p_t^2  +\omega^2 - \frac{3 \s\sqrt{\pi r}\,\rme^{-\frac{t^2}{\pi r+4s}}}{\sqrt{\pi r+4s}}\right]\,.
\ee

Standard semi-classical manipulations show that all the physical effects induced by instantons are encoded in the quantity
\be
\label{om}
\Omega= K \rme^{-S_0[\Phi_{\rm inst}]}\,,
\ee
where
\be\label{ka}
K=\omega\sqrt{\frac{S_0[\Phi_{\rm inst}]}{2\pi}}\left[ \frac{ \omega^2 {\rm det'}\left(\delta^2 S_{\rm E}/\delta\Phi^2\right)_{\Phi_{\rm inst}} }{  {\rm det}\left(\delta^2 S_{\rm E}/\delta\Phi^2\right)_{\pm 1}  }\right]^{-1/2}
\ee
and, as usual, the prime over determinant means omission of the zero mode. The physical meaning of $\Omega$ is that of the resonant frequency between the two minima of the potential due to tunneling (and therefore the tunneling probability per unit of time). Moreover, twice its value is the energy splitting of the even and odd superpositions of the local ground states.

The final step is then the evaluation of the ratio of determinants. To this purpose, the hypothesis of the Gel'fand--Yaglom theorem \cite{GY} are satisfied and we can trade in the calculation of functional determinants for a one-dimensional Cauchy problem.\footnote{See \cite{DK} for a generalization to higher dimensions and \cite{dun07} for an introduction.} 

Since the potential induced by the instanton admits normalizable  bound states, we can apply the Gel'fand--Yaglom theorem on the compact interval $[-T,T]$ setting eventually $T\to \infty$. For this part of the calculation, the non-local problem is very similar to the local one, where the P\"oschl--Teller potential is replaced by a Gaussian one.
The ratio of the determinants is given by the ratio of the  solutions of the corresponding homogeneous differential equations with Cauchy initial conditions, namely
\be\label{y}
\frac{{\rm det'}\left[-\p_t^2  +\omega^2 - \frac{3 \s\sqrt{\pi r}\,\rme^{-\frac{t^2}{\pi r+4s}}}{\sqrt{\pi r+4s}}
   \right]}{\det\left[-\p_t^2 +\omega^2\right]}=\lim_{T\to \infty} \frac{\psi_{\rm inst}(T)}{\psi_{\pm1}(T)}\,,
\ee
where
\ba
\left(\p_t^2 -\omega^2 +\frac{3 \s\sqrt{\pi r}\,\rme^{-\frac{t^2}{\pi r+4s}}}{\sqrt{\pi r+4s}}\right) \psi_{\rm inst}(t)&=&0\,, \quad \psi_{\rm inst}(0)=0\,, \quad \dot\psi_{\rm inst}(0)=1\,,\nonumber\\\label{dif1}\\
\left(\p_t^2  -\omega^2 \right) \psi_{\pm 1}(t)&=&0\,, \quad \psi_{\pm 1}(0)=0\,,  \quad \dot\psi_{\pm 1}(0)=1\,.\nonumber\\\label{dif2}
\ea
The solution of Eq.~\Eq{dif2} is immediate, $\psi_{\pm 1}(t) = \sinh(\omega t)/\omega$, whereas the solution of
Eq.~\Eq{dif1} can be found numerically. To this purpose let us fix the parameters to the values suggested by string theory, Eq.~\Eq{val}, and remember that $m^2=-\s$, so that $\omega^2=2\s=1$.\footnote{This is consistent with the WKB approximation used to define the transmission coefficient, requiring $\omega^3\gg \s/4$; in our case $1>1/8$, which is reasonable.} Then,
\be
K\approx 0.62\,,\qquad \Omega\approx 0.34\,.
\ee
This means that the barrier can be easily penetrated by the non-local system, and that the even superposition of the wave functions localized in the two minima has an energy eigenvalue much lower than that of a particle localized in one of the two minima: If $\psi_{\rm L}$ and $\psi_{\rm R}$ are, respectively, the ground states localized in the left and right minimum, with energy eigenvalue $E_{{\rm L},{\rm R}}=\omega/2$, then the true ground state of the system is $\psi_0=(1/\sqrt{2})(\psi_{\rm L}+ \psi_{\rm R})$, with energy eigenvalue $E_0=\omega/2 -\Omega \approx 0.16\omega$. 

With the same choice of couplings ($-m^2=\s=1/2$), the instanton of the \emph{local} system (Eq.~\Eq{euca} with $r=0$ and $\Phi$ generic) is
\be
\Phi_{\rm local}(t)=\tanh(t/2)\,,
\ee
and one obtains
\be
K_{\rm local}\approx0.65\,,\qquad \Omega_{\rm local}\approx 0.33\,.
\ee
Thus, concerning the tunneling frequency, the difference between the local and non-local cases is about $3\%$. Taking into account the level of approximation of the scaling formula and its propagation (that does not exceed $2\%$ in the model under investigation) one can conclude that non-locality does affect tunneling events, although in a very mild way.
These results are also compatible with the difference (in $L^2$ norm) between the local and non-local instanton solutions, $\tanh (t/2)$ and $\Phi_{\rm inst}(z)$: although small, it is well above $2\%$. Consequently, there is a quantitative departure from local physics. Conversely, it is remarkable that all the effect of non-locality is just a very mild modification of the physical parameters, especially in view of the drastic changes that non-locality provides, instead, in the Minkowskian classical solutions of the same problem.


\subsection{Brane tension}\label{tension}

In order to complete the proof that Eq.~\Eq{erf} is a solution of OSFT, we must verify that it describes a BPS brane configuration. We refer to \cite{sen04} for a review on the subject.

According to Sen's first and second `conjectures' \cite{Sen99,Se992,Hor98,Se98a,Se98b,KMM1},\footnote{As a point of fact these are no longer conjectures, since they have been proven or strongly supported numerically in the level truncation scheme \cite{HK,BSZ,IN,MSZ,RSZ1} and analytically \cite{HKM,KMM2,Oku02,Oka02,CFGNO,Sch05}. See \cite{FK} for an updated discussion on the subject and references.} the tachyon vacuum corresponds to an unstable (non-BPS) D$p$-brane (whose volume ${\cal V}_{p+1}$ we normalize to 1), which decays into a stable configuration. Therefore, at its local maximum the effective tachyon potential equals the tension of the non-BPS D$p$-brane, which is \cite{Sen99} 
\be\label{TDp}
{\cal T}_p=\frac{1}{2\pi^2g_o^2}\,,
\ee
where $g_o$ is the open string coupling. When $p=9$, the brane coincides with the target spacetime of Type I/IIA theory.
On the other hand, a tachyonic kink solution $\Phi(x)$ interpolating between the two minima is interpreted as the BPS D$(p-1)$-brane, localized at $x=0$, into which the unstable D$p$-brane has decayed \cite{Hor98,Se98a,Se98b}. The tension of this brane is
\be
\tilde{\cal T}_{p-1}=\sqrt{2}\pi{\cal T}_p.
\ee
The prefactor takes into account reduction of dimensionality of the brane (${\cal T}_{p-1}=2\pi{\cal T}_p$) and the fact that the tension of an unstable D$p$-brane is $\sqrt{2}$ times the tension of a BPS D$p$-brane.

If $t=x$ is regarded as a space coordinate in the $(p+1)$-dimensional target spacetime, our solution (which is now a Lorentzian soliton) is a candidate realizing the above picture. To show this, we must first revert to the original effective action of string field theory (with approximated local potential) and then fix the normalization of the solution. The former is 
\be\label{eucal}
S_*=\frac1{2g_o^2}\int \rmd^{p+1}x \left[\tilde\Phi\left(\p_x^2+\frac12\right)\tilde\Phi-\frac{\s}2\Phi^4- 2\Lambda\right]\,,
\ee
where $\Lambda$ is a constant and
\be
\tilde\Phi=\rme^{-s\p_x^2}\Phi=\Phi_0\erf\left(\frac{x}{2}\right)\,.
\ee
Here we used $r=\k^2$, $r-s=1$. In the previous subsections we chose the normalization $\Phi_0=1$. Otherwise, the quartic term is rescaled as $1/2=\s\to\s/\Phi_0^2=1/(2\Phi_0^2)$. Now we can fix this residual free parameter so that $V(0)-V(\pm\Phi_0)=V(0)={\cal T}_p$, yielding
\be\label{fix}
\Phi_0^2=8\Lambda=8g_o^2{\cal T}_p\,.
\ee
The truncation level of the action affects the value the non-BPS brane tension ${\cal T}_p$ and possibly the ratio
\be\label{ratio}
-\frac{S_*[\tilde\Phi,\Phi]}{{\cal T}_p} \stackrel{?}{=} \sqrt{2}\pi\,.
\ee
A non-trivial check to perform on the thus-established solution is to see whether Eq.~\Eq{ratio} is satisfied.  Explicitly, and using the second duplication formula,
\ba
-\frac{S_*[\tilde\Phi,\Phi]}{{\cal T}_p}&\approx& \frac{8}{\sqrt{2\pi}}
+\pi\left(2-\frac{3\k}{2^{5/4}}\right)\nonumber\\
&\approx& 4.558,\label{xxx}
\ea
where $\k=\k(3,1/2)$. This value reproduces the expected tension ratio $\sqrt{2}\pi\approx 4.443$ within $2.6\%$. Taking instead the string value Eq.~\Eq{val}, one gets $-S_*/{\cal T}_p\approx 4.584$, $103.2\%$ the theoretical ratio. Evaluating the action exactly, i.e., without using the duplication formul\ae, one obtains 
\be\label{TTt}
-\frac{S_*[\tilde\Phi,\Phi]}{{\cal T}_p}\approx 4.435 = 0.998\times (\sqrt{2}\pi)
\ee
for $\k=\k(3,1/2)$, and $-S_*/{\cal T}_p\approx 4.461\approx 1.004\times (\sqrt{2}\pi)$ for Eq.~\Eq{val}. 

Considering that $\erf(x/2)$ was regarded as the approximate solution of the lowest-level approximate effective action, the agreement is impressive.


\section{Supersymmetric OSFT: Non-local potential}\label{exsu}

We now turn to the real effective $(1/2,1)$-level-truncated potential of OSFT which contains `nested' non-local operators. In particular, we would like to reproduce Eq.~\Eq{exa},
\[
(\p_t^2-m^2)\rme^{-s\p_t^2}\Phi=\s\Phi\,\rme^{s\p_t^2}\Phi^2\,,
\]
for some constants $m^2$, $s$ and $\s$. Since the potential itself is dressed with a non-local operator, for purely technical reasons this system has been typically traded for its simplified form Eq.~\Eq{geom} in the literature. However, we are in a position to bypass this approximation and deal with the original case in a rather simple fashion.

We can appreciate the advantage of using the duplication formul\ae\ when dealing with equations like \Eq{exa}. By construction, all the $\Phi$ functions satisfy the heat equation and the action of the non-local operators is a translation in the evolution parameter $r$, Eq.~\Eq{tra}. However, products of the scalar field generally do not satisfy the heat equation and therefore Eq.~\Eq{tra} cannot be used. The role of the duplication formul\ae\ is then clear: They select the best approximation for non-local products of $\Phi$ functions still satisfying the heat equation. Consequently, we shall reduce Eq.~\Eq{exa} to an algebraic equation, as was done also in \cite{cuta2} for cosmological models.

The analysis of the previous section taught us that non-local effects are rather mild and the dynamics does not change much with respect to the local system. Therefore we do not expect the non-local operator on the left-hand side of  Eq.~\Eq{exa} to modify qualitatively the behaviour of the solution with local potential. 

Indeed, an approximate solution of Eq.~\Eq{exa} is again the error function $\Phi(1/2,z)$. 
Eq.~\Eq{use} for $\a=1/2$ can be written as
\be\label{use2}
\p_z^2\Phi\left(\frac12, z\right)=2\left[\Phi\left(\frac32, z\right)-\Phi\left(\frac12, z\right)\right]\,.
\ee
On the other hand, according to Eq.~\Eq{tra} the non-local operator can be recast as
\be\label{difz}
\rme^{\b\p^2_z}\Phi\left(\frac12,z\right)=\Phi\left(\frac12,\frac{z}{\sqrt{1+4\b}}\right)\,.
\ee
Let
$m^2=-1/2$ and
\be
s=\frac{\k_*^2-1}{\k^2}r\,,
\ee
where $\k_*$ and $r$ are constants later to be determined. The right-hand side of Eq.~\Eq{exa} is $\s$ times
\ba
\Phi\left(\frac12,z\right)\rme^{\frac{s}{4r}\p_z^2}\Phi^2\left(\frac12,z\right) &\ \stackrel{\Eq{dupli2}}{\approx}\ & \Phi\left(\frac12,z\right)\rme^{\frac{s}{4r}\p_z^2}\Phi(1,\k z)\nonumber\\
&\ \stackrel{\Eq{gf1}}{=}\ &\Phi\left(\frac12,z\right) \left[1-\frac1{\k_*}\rme^{-\left(\frac{\k }{\k_*}z\right)^2}\right]\,,\label{use4}
\ea
while the left-hand side is $1/(2r)$ times
\ba
\left(\frac12\p_z^2+r\right)\rme^{-\frac{s}{4r}\p_z^2}\Phi\left(\frac12,z\right)&\ \stackrel{\Eq{difz}}{=}\ &\tk^2\left(\frac12\p_{\tk z}^2+\frac{r}{\tk^2}\right)\Phi\left(\frac12, \tk z\right)\nonumber\\
&\ \stackrel{\Eq{use2}}{=}\ & 
\tk^2\left[\Phi\left(\frac32, \tk z\right)+\left(\frac{r}{\tk^2}-1\right)\Phi\left(\frac12, \tk z\right)\right],\nonumber\\\label{use3}
\ea
where 
\be
\tk\equiv \sqrt{\frac{\k^2}{1+\k^2-\k_*^2}}=\sqrt{\frac{r}{r-s}}.
\ee
Incidentally, notice that a plot of $\Phi^3$ against the right-hand side of Eq.~\Eq{use4} clearly shows that the maximum deviation from the local and the non-local case is at and near the origin, a fact which had been already realized in the comparison between local and non-local kinks with local potential \cite{MZ,are04}.

Combining Eqs.~\Eq{use4} and \Eq{use3} one obtains the algebraic equation
\be
\Phi\left(\frac32, \tk z\right)+\frac{2r\s}{\k_*\tk^2}\Phi\left(\frac12,z\right)\rme^{-\frac{\k^2}{\k_*^2}z^2}
\approx \frac{2r\s}{\tk^2}\Phi\left(\frac12,z\right)-\left(\frac{r}{\tk^2}-1\right)\Phi\left(\frac12, \tk z\right)\,.\label{alge}
\ee
At large $z$ the correct asymptotic behaviour is obtained when $\s=1/2$. The simplest possible case is given by
\be\label{kkk}
\k_*=\tk=\k=\frac{2}{\sqrt{\pi}}\,.
\ee
Then, one can check that
\be\label{solu}
\Delta_{\rm max}\approx 0.6\%\,,\qquad \delta\approx 0.7\%\,,\qquad {\rm for}\qquad r\approx 1.418\,,
\ee
which is well below the error of Eq.~\Eq{dupli2}. If one optimizes also for $\k$ or employs the first duplication formula \Eq{dupli1} in the left-hand side of Eq.~\Eq{alge} with $\bk$ given by Eq.~\Eq{bk}, $\Delta_{\rm max}$ is about the same. Amusingly, if instead one takes the isotropic value $\bk=\k$, the maximum error drops below $0.04\%$! At any rate, the global accuracy of the solution is effectively determined only by the one of the duplication formula.

The exponent $s$ associated with Eq.~\Eq{solu} is $s\approx 0.304$, which is about $40\%$ off the string value Eq.~\Eq{stv}. This should not discourage the reader for two reasons. First, it is not difficult to see that Eq.~\Eq{erf} is still a good solution of the system when Eq.~\Eq{stv} is enforced. In that case, $\k_*=\sqrt{1+\k^2 s_*/r}$ and the integral in Eq.~\Eq{del} is minimized with respect to $r$ when $r \approx 1.620$, for which $\Delta_{\rm max} \approx 4.7\%$. One can refine the approximation by letting $\k$ free and fix the pair $\k,r$ so that the error on the second duplication formula for $(n,\a)=(2,1/2)$ and the one on Eq.~\Eq{alge} are the same. The result is
$\k \approx 1.172$, $r\approx 1.616$, $\Delta_{\rm max} \approx 3.8\%$. To improve this solution one could deform it locally, for instance as done in \cite{vla05}, with the requirement that the perturbed solution still obeys the diffusion equation. We shall not pursue this possibility here.

Secondly, the brane tension ratio is about the same as, if not better than, the value obtained in the previous section. Take the Lorentzian string action with parameters given in Eqs.~\Eq{kkk} and \Eq{solu},
\be\label{eucal2}
S_*=\frac1{2g_o^2}\int \rmd^{p+1}x \left[\tilde\Phi\left(\p_x^2+\frac12\right)\tilde\Phi-\frac{\s}2\left(\rme^{s\p_x^2}\Phi^2\right)^2- 2\Lambda\right]\,,
\ee
where now $r-s\neq 1$. Using the second duplication formula one obtains
\ba
-\frac{S_*[\tilde\Phi,\Phi]}{{\cal T}_p}&\approx& \frac{8}{\sqrt{2\pi (r-s)}}
+\pi\left[2\sqrt{r-s}-2\sqrt{r}+\frac{\sqrt{\pi}\,r}{\sqrt{2(\pi r+4s})}\right]
\nonumber\\
&\approx& 4.517 = 1.017\times (\sqrt{2}\pi).
\ea
We conjecture that the exact result is extremely close to the theoretical value, although one should employ a careful numerical procedure (for instance the iterative method of \cite{Jo081,Jo082}) to evaluate the fully non-local action.


\section{Conclusions}\label{conc}

We can summarize the main achievements as follows:
\begin{itemize}
\item We have derived approximate duplication formul\ae\ which express products and powers of incomplete gamma functions as a single $\g$ with rescaled arguments. Their level of approximation (about $1\%$) is under control and sufficient to easily construct instantonic tachyon solutions of supersymmetric string field theory, where both the effective kinetic term and potential are highly non-local.
\item The solution of the Euclidean OSFT problem at lowest-order truncation level with \emph{local} potential is the error function $\erf(z)$. Despite the non-locality of the tachyon kinetic term has been fully taken into account, with respect to the `local' instanton there is little quantitative and no qualitative difference in the physics of tunneling events.
\item The solution of the exact problem is still the error function, with the same global error. With respect to the solution for the local potential, the only difference is in the value of the parameters in the equation of motion.
\item To the best of our knowledge, \emph{this is the first time a solution of the non-local effective equation \Eq{exa} is found either analytically or numerically}.
\item Interpreted as a spatial kink, the solution in both versions of the theory is a Lorentzian soliton and, according to Sen's `second conjecture', describes the decay of a non-BPS D$9$-brane into a codimension-1 BPS brane. This interpretation is fully confirmed by the brane tension ratio we obtained, which is in excellent agreement ($99.8\%$) with the expected value. The role of non-locality was essential in the recovery of this result.\footnote{For instance, taking the formal `local' limit $\k\to 1$ ($s=0$) in Eq.~\Eq{xxx} one obtains $62\%$ of the brane tension ratio.}
\end{itemize}

Concerning the last point, the importance of \emph{lowest-level} and \emph{non-local} solutions has become apparent. Due to technical difficulties, off-shell calculations have been carried out only at lowest truncation level for both the bosonic and supersymmetric string, so that the main effort has been focussed on the effective spacetime action with non-local operators for the tachyon field only. This is sufficient to describe non-perturbative processes, as the tachyon is the only particle field with negative mass, and it is expected to encode the physics of unstable geometric configurations decaying into different others. Higher-level calculations as well as the recent analytic solutions in the full theory have confirmed that the lowest-level tachyon dynamics captures most of the features of brane decays. Eq.~\Eq{TTt} confirms this statement but adds to it the desirability of taking non-locality fully into account: Lowest-order calculations which are on-shell (i.e., local) or quasi-local (i.e., truncating non-local operators) are unable to achieve this degree of accuracy.

We also notice that, for us, the tension of the single branes is unknown and is actually fixed by hand, Eq.~\Eq{fix}. The ratio of brane tensions, however, is close to the expected value. In other words, non-locality can tell the difference between brane configurations with very good accuracy, but is insensitive of the properties of the individual branes. This is expected, as non-local operators govern the relative value of fields $\Phi(r,x)$ and $\Phi(r',x)$ in the equations of motion, but it does not affect the local structure of these equations, which is given by the level approximation. This is why the diffusion equation method plays an important role in the descriptions of the non-perturbative physics of very different in- and out-states. Its applications could be extended to other geometries (such as lumps of different codimension), but only after a spacetime non-local effective action has been constructed. This latter step is often very non-trivial and must be done with traditional techniques but, once this is accomplished, the rest would be under control.\footnote{Large-codimension lumps are expected to be described with a lesser degree of accuracy. Take the simplest case of a factorizable action (a typical example is the Gaussian case). The smallest error is obtained for codimension 1, because the product of more elements would propagate the error.}

One of the most remarkable consequence of our procedure is the recovery `out of the blue' of the parameters and the physics of OSFT, Eqs.~\Eq{stv} and \Eq{TTt}, for the supersymmetric local potential. We believe that the rather small difference between the actual values and those of OSFT is due to the approximation intrinsic of the duplication formula. This result is highly non-trivial. The mass and non-local exponent appear as separate inputs in the effective equation of the OSFT tachyon, although both are determined by conformal invariance. Obviously, a time rescaling can change their ratio, which is precisely the job done by the parameter $r$ in our model. However, their product $m^2s=(\kappa^2-1)/2$ is independent of $r$ and is determined only by the value of $\kappa$, which comes from the scaling formula for a particular type of potential. It attracts all our attention that the effective equation of the string tachyon with the same values of the coupling constants, as well as the brane descent relation in Sen's tachyon condensation, have been obtained starting from an apparently different framework.

In other words, one can pose the question: What is the scalar field theory which describes tunneling between two inequivalent vacua and whose solutions obey the diffusion equation? The answer, surprisingly, is: The Euclidean tachyonic sector of open string field theory. The fact that string field theory may be viewed as a diffusing system was already pointed out in \cite{roll,cuta4}, where tachyon solutions of OSFT and boundary string field theory were mapped onto each other. We will formalize this picture in a forthcoming study and show that the diffusion equation implements conformal invariance at the level of the effective dynamics.


\ack
G.C. is supported by NSF Grant No. PHY0854743, the George A. and Margaret M. Downsbrough Endowment and the Eberly research funds of Penn State. G.N.\ is partly supported by INFN of Italy.
 

\appendix

\section{Derivation of Eq.~\Eq{exa} from the OSFT action}

Contrary to the bosonic string, there are several proposals for superstring field theory, the first being Witten's \cite{wi86b,con1,con2,con3,con4,DR}. The action was later modified by \cite{PTY,AMZ1,AMZ2}. On a single non-BPS D$p$-brane, 
\ba
S&=&-\frac1{g_o^2}\int Y_{-2}\left(\frac1{2\alpha'}\Psi_+* Q\Psi_++\frac13\Psi_+*\Psi_+*\Psi_+\right.\nonumber\\
&&\qquad\qquad\qquad\left.+\frac1{2\alpha'}\Psi_-* Q\Psi_--\Psi_+*\Psi_-*\Psi_-\right),
\ea
where $g_o$ is the open string coupling constant, $\int$ is the path integral over matter and ghost fields, $Y_{-2}$ is a double-step inverse picture-changing operator, $\alpha'$ is the Regge slope, $Q$ is the BRST operator, * is a noncommutative product, and the string field $\Psi_\pm$ is a linear superposition of states (made of matter (super)fields $X^\mu$, $\psi^\mu$ and (super)ghosts $b$, $c$, $\b$, $\gamma$) in the GSO$(\pm)$ sectors \cite{Se99c}, respectively. The coefficients of these states correspond to the particle fields $f_i$ of the string spectrum.

The operator $Y_{-2}$ can be either chiral and local \cite{AMZ1,AMZ2} or nonchiral and bilocal \cite{PTY} (see the literature and the review \cite{ohm01} for full details). These two theories predict the same tree-level on-shell amplitudes but different off-shell sectors.
From now one we concentrate on the nonchiral version \cite{PTY,AKBM,AJK} and set $\alpha'=1$. 

Expanding the string field in an $L_0$ eigenbasis, let $h_i$ be the conformal dimension of the vertex operator associated with $f_i$. One defines the \emph{level} of $f_i$ to be $M=h_i+1$. By definition the tachyon has level $1/2$. The \emph{truncation level} $(M,N)$ selects all quadratic and cubic terms of total level no greater than $N$ made of fields with level $\leq M$ \cite{MSZ,SZ}; in general, $2M\leq N\leq 3M$ \cite{AKBM}.

In the 0 picture and at level $(1/2,1)$, which is the lowest for the susy tachyon effective action, all particle fields in $\Psi_\pm$ are neglected except the tachyonic one, labeled $\tilde\Phi(x)$ and depending on the center-of-mass coordinate $x$ of the string, and an auxiliary level $-1$ field $u(x)$. The Fock-space expansion of the string field is truncated so that $\Psi_+(z) =\int \rmd^{p+1} k\, u(k) c(z)\rme^{\rmi k\cdot X(z)}$ and $\Psi_-(z) =\int \rmd^{p+1} k\,\tilde\Phi(k) \gamma(z)\rme^{\rmi k\cdot X(z)}$. The spacetime action reads \cite{AKBM,AJK}
\be
S=\frac1{g_o^2}\int \rmd^{p+1}x\left[\frac12 \tilde\Phi\B\tilde\Phi+\frac14\tilde\Phi^2+u^2-\frac{\rme^{s_*}}3(\rme^{s_*\B/2}u)\Phi^2\right]\,,\label{su}
\ee
where $s_*$ is given in Eq.~\Eq{stv}, $\B=\p_\mu \p^\mu$ and $\Phi\equiv\rme^{s_*\B/2}\tilde\Phi$. Combining the equations of motion for $u$ and $\Phi$,
\ba
&&u - \frac{\rme^{s_*}}{6}\rme^{s_*\B/2}\Phi^2=0\,,\\
&& \left(\B+\frac12\right)\rme^{-s_*\B}\Phi- \frac{2\rme^{s_*}}{3}\Phi \rme^{s_*\B/2}u=0\,,
\ea
one obtains an equation of motion for $\Phi$ alone \cite{AJK}. It is the presence of the field $u$ in the cubic action Eq.~\Eq{su} which generates a quartic term in the tachyon effective action.

Specializing to a one-dimensional spatial configuration (or a Wick-rotated homogeneous background), Eq.~\Eq{eucal2} (respectively, Eq.~\Eq{exa}) is recovered for specific values of $m^2$, $s$ and $\s$.


\end{document}